\documentclass[reprint, amsmath,amssymb, aps]{revtex4-1}

\usepackage{graphicx}
\usepackage{epstopdf}
\usepackage{dcolumn}
\usepackage{bm}

\begin{document}

\preprint{APS/123-QED}

\title{Do dipole correlations exist in two-dimensional spin ice?}

\author{M. I. Ryzhkin}
 \email{ryzhkin@bk.ru}
\affiliation{ Institute of Solid State Physics of Russian Academy of Sciences\\
 2 Academician Ossipyan str., Chernogolovka Moscow District, Russia, 142432}

\date{\today}

\begin{abstract}
Here we study the statistical properties of two-dimensional spin ice in its ground state by the Monte Carlo simulation method. Using a new sampling algorithm, we show that the short-range ice rule in two dimensions gives rise to long-range but not dipole-like correlations, to non-Gaussian probability density function for magnetization and to non-extensive conditional entropy (entropy with given value of magnetization). 
\begin{description}
\item[PACS numbers]
75.10.Hk, 75.40.Mg, 05.10.Ln
\end{description}
\end{abstract}

\maketitle

{\sl Introduction.} - Originally, the name "spin ice" was given to compounds like $\rm Ho_2Ti_2O_7, Dy_2Ti_2O_7$ which demonstrate unusual magnetic ordering \cite{Harris1997}.  The magnetic ions in these compounds sit at the vertices of regular tetrahedrons linked into a three dimensional pyrochlore lattice. Owing to strong anisotropy, the spins of magnetic ions can be directed along local anisotropic axes connecting centers of nearest tetrahedrons. The ground state is characterized by the ice rule: two spins of each tetrahedron are directed toward its center, and two other spins, from its center. Subsequently, the term spin ice was applied to any physical system with spins oriented according to the ice rule (two spins in and two spins out rule). The spin ice shows such amazing properties as an exponentially degenerate ground state \cite{Ramirez1999}, long-range dipole correlations \cite{Youngblood1980, Huse2003, Isakov2004, Henley2005}, self-screening of dipole interaction \cite{Gingras2001, Isakov2005} and emergent magnetic monopoles \cite{Ryzhkin2005, Castelnovo2008}. It is also considered as a classical analogue of quantum spin liquid \cite{Balents2010}. 

The two-dimensional implementation of spin ice called an artificial spin ice creates new possibilities for information processing technology \cite{Wang2006, Kitaev2003}. The corresponding theoretical model belongs to the class of six-vertex models allowing exact solutions (for some specific values of parameters) \cite{Lieb1967, Sutherland1968, Baxter1982}. The availability of analytical and experimental results makes two-dimensional spin ice a unique testing area for debugging of numerical algorithms. That explains our initial purpose: to develop a new sampling algorithm for the Monte Carlo simulation of spin ice using the two-dimensional spin ice model. However, the obtained results were so unexpected that the study of statistical properties of two-dimensional spin ice became the main substance of the paper. 

Let us describe the initial and modified purpose in detail. The model of two-dimensional spin ice can be described as follows. Ising spins sit at bonds of square lattice and are only allowed to be directed along the bonds (see Fig. 1). All vertices satisfying ice rule (two spins in and two spins out of each vertex) have the same statistical weight or energy (taken as a zero), violations of ice rule give some finite positive energies.  Therefore, in ground state, there are only correct vertices which satisfy ice rule (see Fig.1). It is easy to see that in a ground state, the flip of a spin results in violations of ice rule for two vertices: with three (one) spins in and one (three) spin out from the vertex. These vertices carry an effective magnetic charge and have finite creation energy. By this reason, a usual sampling algorithm (random choice of spin and its flip with the probability defined by Metropolis function) becomes ineffective at low temperatures since the simulation time grows exponentially as the temperature decreases  \cite{Binder2010}. Moreover, it is not applicable for the studies of the ground state at all. 

\begin{figure}
\includegraphics[width=0.4\textwidth]{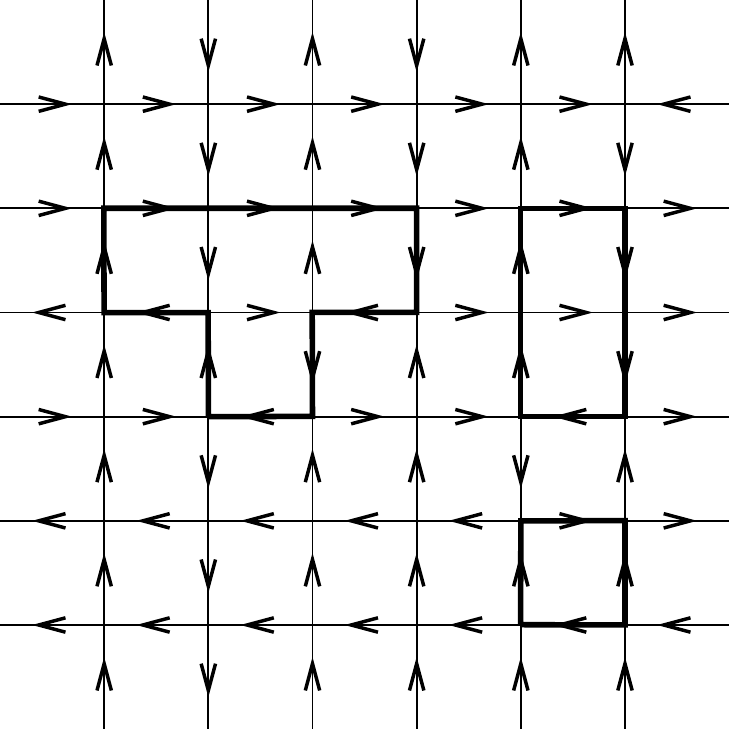}
\caption{Ising spins in two-dimensional spin ice sit on each bond of square lattice and are constrained to point in only toward or out of the nearest neighbor vertices. In the ground state configurations, all vertices satisfying the ice rule (two spins in and two spins out of each vertex) have the same energy. The loops of spins aligned along loops are highlighted in bold line.}
\label{fig.1}
\end{figure}

To overcome this problem, special algorithms were developed. They are all based on searching for a closed path of spins aligned along the path (see, as an example, the loop in figure 1) and on the following collective flip of all spins in the loop \cite{Rahman1972, Yanagawa1979, Adams1984, Barkema1998}. Such collective flip do not violate ice rule and strongly reduce the simulation time. But they do not solve the problem completely. In fact, the configuration space of spin ices has a very involved structure splitting into separated topological sectors \cite{Jaubert2013}. The transitions between the sectors are kinetically slowed down, in particular, they cannot be realizable by loop flips. Indeed, loop flips do not change magnetization and, therefore, they cannot lead to transitions between topological sectors with different magnetization. This is the main reason why we planned to explore a new sampling algorithm which could provide spin configurations from different topological sectors. 

It is generally believed, that the simple two-dimensional model described above keeps all essential features of three-dimensional spin ice, for example such properties as (1) dipole correlations, (2) extensive conditional (an entropy at given magnetization) entropy and (3) Gaussian density distribution function of magnetization. However, our results reveal that in two dimensions conditional entropy is non-extensive, probability density function of magnetization is non-Gaussian, and these special features are consequences of long-range non-dipole correlations.

{\sl Method.} - A specific feature of our approach is a new sampling algorithm consisting of two stages. At the first stage, we define absolutely by chance all spins of cluster. As a result, we get an absolutely random configuration with a great number of incorrect vertices, where ice rule is broken. At the second stage, we use the Metropolis  annealing algorithm to reduce the number of incorrect vertices to zero. This means that we take a random spin and try to turn it with the probability defined by the Metropolis function \cite{Binder2010}. This procedure is repeated until a ground state configuration is obtained. 

After the second stage, we get a ground state configuration which can be used for calculation of physical variables. We repeat the two stages procedure many times to get settled and consistent results. This sampling algorithm creates the configurations distributed over configurational space more uniformly than the usual "loop procedure" \cite{Rahman1972, Yanagawa1979, Adams1984, Barkema1998}. The computer implementation of our algorithm is simple, rather fast and it can be directly modified for many processor systems by a simple accumulation of results from individual processor.

With this algorithm, we generate $T$ ground state configurations for square clusters of two-dimensional spin ice with $N'$  spins. We use free boundary conditions and exclude several outer layers from further computations to avoid boundary effects. For each configuration, we calculate magnetization according to the formula
\begin{equation}
{\bf M}(t)=\sum_{k=1}^{N}\sigma_k (t){\bf e}_k
\label{eq:1}
\end{equation}
Here $\sigma_{k}(t)=\pm1$ is the spin on the bond $k$ at the configuration with integer index $t$,  ${\bf e}_k$ are unit vectors equal  to ${\bf e}_x, {\bf e}_y$ for the horizontal and vertical bonds respectively, and summation is over all $N<N'$ interior spins of the cluster. Note that $N<N'$ because we exclude two outer layers of vertices. The numbers of spins in the interior domain are taken from $N=40$ to $N=77224$, and the number of configuration is taken about $T=10^5$. In individual cases, we take $N=1986024$ and $T=10^7$. From the results of simulation, we calculate one and two dimensional histograms (or sampling probability density functions) $g(M_x),G(M_x,M_y)$, and the first sampling moments
\begin{align}
&E(M_{x,y}^m)=T^{-1}\sum_{t=1}^{T}M_{x,y}^m(t),\label{eq:2}\\
&E(M_{x,y}^mM_{x,y}^n)=T^{-1}\sum_{t=1}^{T}M_{x,y}^m(t)M_{x,y}^n(t),\label{eq:3}
\end{align}
for $m,n=1,2,3,4$. All quantities are calculated as functions of $N$, and for future analysis we also calculate them for the non-interacting model whose properties are well known.

\begin{figure}
\includegraphics[width=0.5\textwidth]{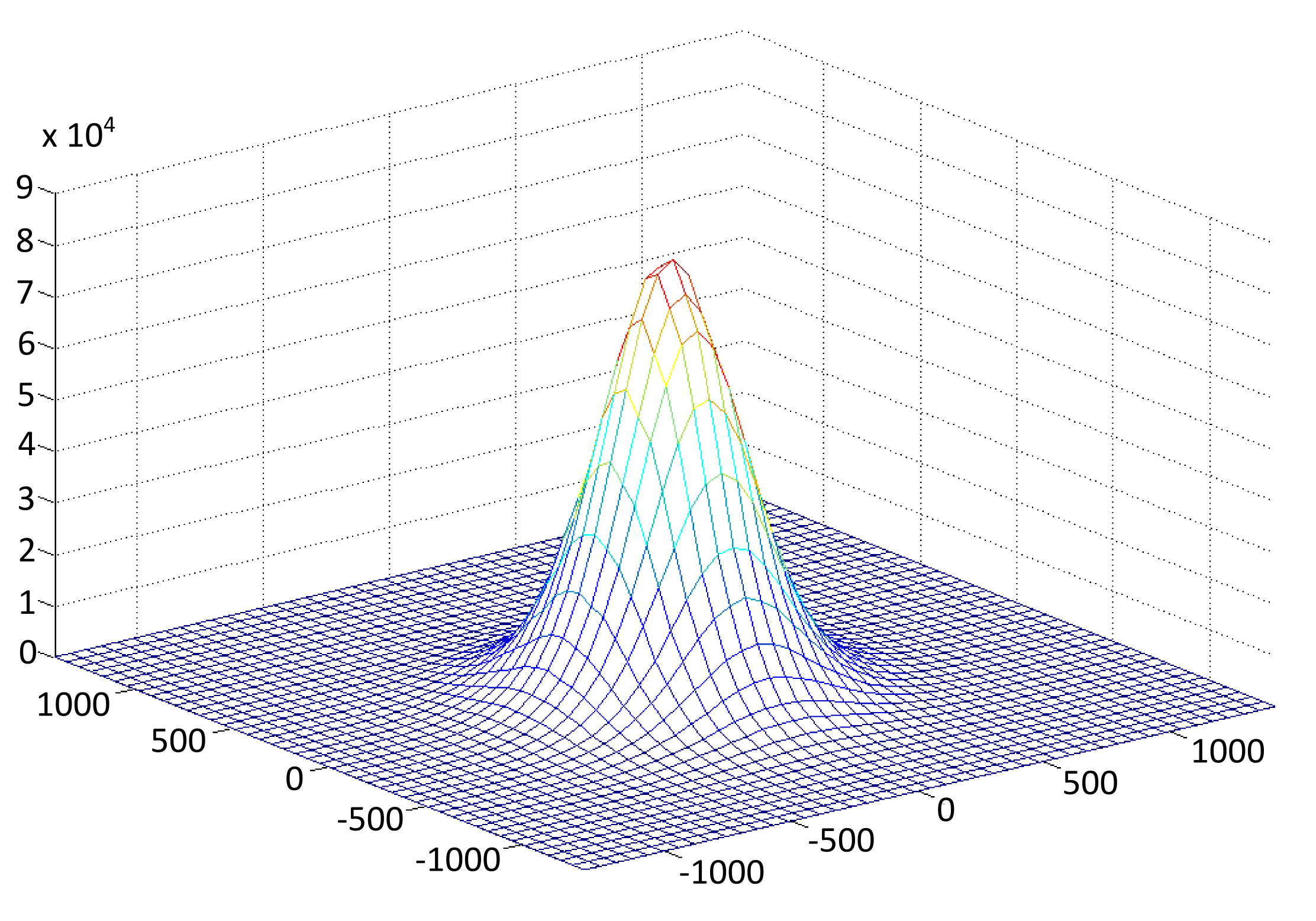}
\caption{Two-dimensional histogram $G(M_x,M_y)$ for the cluster of $1986024$ spins as a function of magnetization. Total number of configurations is about $10^7$.}
\label{fig.2}
\end{figure}

{\sl Results} - First, we have found that simple generators of pseudo-random numbers result in artificial effects and correlations. To test the employed pseudo-random number generators, we compared the simulation performed with them and the one made with files of genuine random numbers \cite{Marsaglia1995}. Unfortunately, one could not use only these files to perform long Monte Carlo simulation, because they are rather short. We have found that  the pseudo-random number generator called Mersenne twister gives a sufficiently good results, moreover it has an enormous period of $2^{19937}$ that allows very long simulations \cite{Matsumoto2007}.

Our results are given in Figs.2-4. In Fig.2, we give a two-dimensional histogram $G(M_x,M_y)$ as a function of magnetization. The calculations of the first sampling moments show that odd moments equal zero $E(M_{x,y}^{1,3})=0$, and the even ones are isotropic $E(M_{x}^{2,4})=E(M_{y}^{2,4})$, in addition the mixed moment $E(M_{x}M_{y})=0$. The accuracy of these statements becomes higher as the simulation time $T$ grows (that has been checked up to $T=10^7$). In Fig.3, we give the dispersions of magnetization as a function of the number of spins $N$ for both the spin ice and for the system without any interaction between spins. In Fig.4, the dependence of  $\partial{\ln{G(M)}}/\partial M^2$ upon $N$ is also shown for both the spin ice and for a paramagnetic system. Now let us discuss the obtained results.

\begin{figure}
\includegraphics[width=0.5\textwidth]{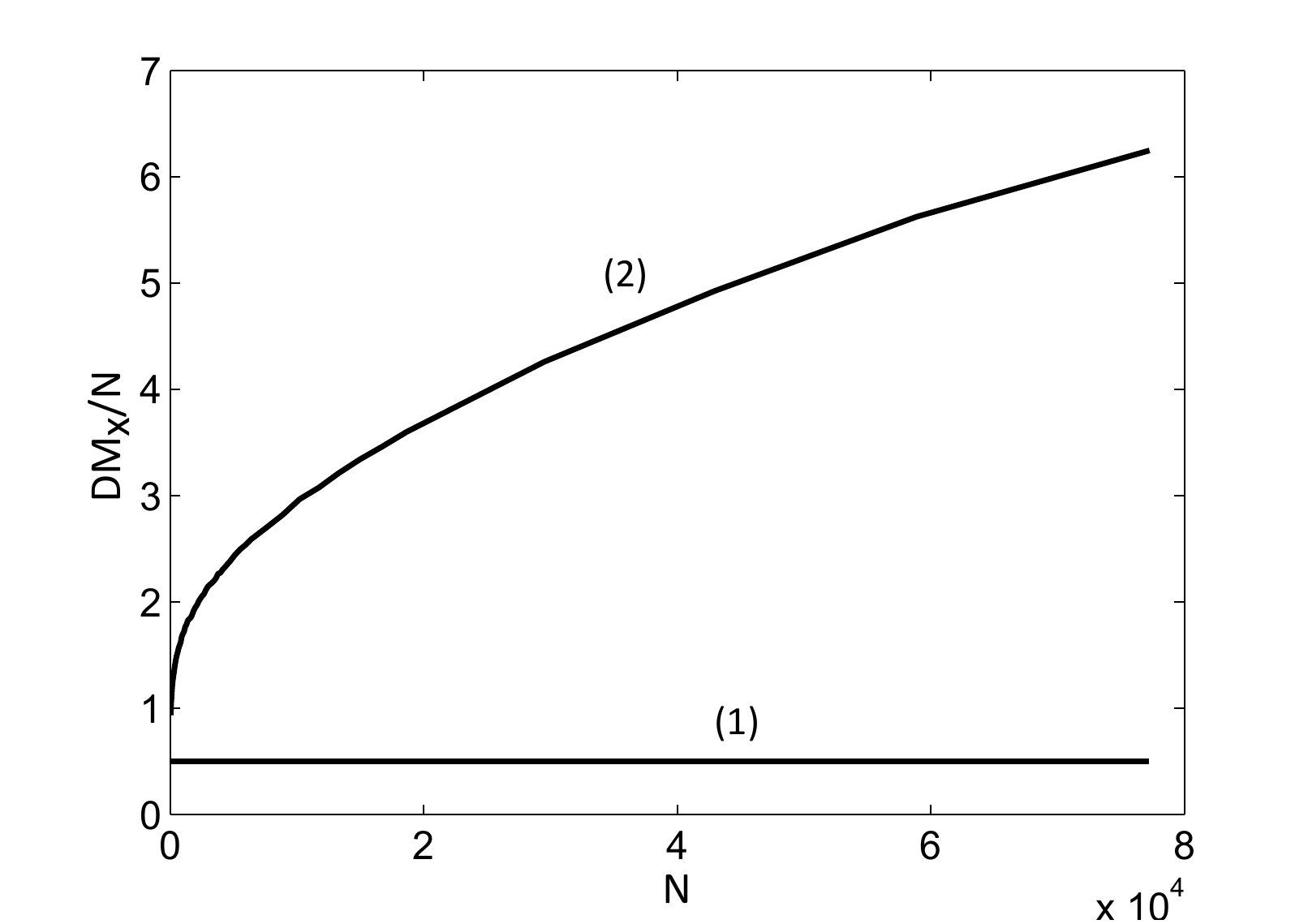}
\caption{Dispersions of magnetization as function of the number of spins for the non-interacting spins (1) and for the spin ice (2).}
\label{fig.3}
\end{figure}

{\sl Discussion} - Fig.2 reveals that a density of configuration is isotropic and has a peak at zero magnetization ${\bf M}=0$. These results were expected from general considerations, and can be regarded as a proof of the absence of essential errors in the simulation procedure. Indeed, the plot in Fig.2 is very sensitive to any attempts to accelerate the configuration generation algorithm via insertion of some faster but determinate elements. For example, replacement of the random choice of spins for possible flip by sequential scan of all spins results in a shift of the maximum from zero.  Further,  it becomes clear from the histogram shape that $E(M_{x,y}^{1,3})=0$ and $E(M_{x}^{2,4})=E(M_{y}^{2,4})$, which was also tested by direct calculations. Accordingly, the histogram peak for the noninteracting system also locates at ${\bf M}=0$, possesses axial symmetry, but exhibits a more narrow peak. 

From Fig.3 one can see that for a noninteracting system the quantity $D(M_x)/N=const$ as should be expected for noninteracting and, therefore, for non-correlated spins. This result is in agreement with central limiting theorem for non-correlated spins. But for spin ice with a short-range interaction determined by the ice rule, the quantity $D(M_x)/N \neq const$. We have found that the dependence of $D(M_x)/N$ upon $N$ can be approximated by power-like function
\begin{equation}
D(M_x)/N=a+bN^{\alpha}
\label{eq:4}
\end{equation}
where $a,b$ are some positive constants and $\alpha = 0.37$. This important result implies violation of the central limit theorem for spins interacting according to the ice rule. Therefore, it is natural to assume that the probability density function for magnetization can differ from the Gaussian one.

This assumption will be thoroughly examined and confirmed below. However, it is first reasonable to consider the relation between Eq.(4) and assertion about the dipolar character of the correlation function  \cite{Youngblood1980, Huse2003, Isakov2004, Henley2005}. The latter, strictly speaking, concerns the spin ice mode in three dimensions. But its extension for two dimensions is straightforward and quite common. In our case, the above statement can be written in the form:
\begin{equation}
S_{\alpha \beta}({\bf r})=\Big\langle M_{\alpha}({\bf r})M_{\beta}(0)\Big\rangle\propto\frac{\delta_{\alpha\beta}-2n_{\alpha}n_{\beta}}{r^2}
\label{eq:5}
\end{equation}
where  $n_{\alpha}=r_{\alpha}/r$.  Let us show that the result expressed by Eq. (4) disagrees with that of Eq.(5). To this end we estimate the variance (below summation is restricted to the horizontal bonds):
\begin{eqnarray}
DM_x=E\Bigg[\Bigg(\sum_{i=1}^{N/2}\sigma_{i}\Bigg)\Bigg(\sum_{k=1}^{N/2}\sigma_{k}\Bigg)\Bigg]=\qquad \qquad\\
E\sum_{i=1}^{N/2}\sigma_{i}^{2}+E\sum_{i \ne k}^{N/2}\sigma_i\sigma_k
\leq \frac{N}{2}\Big[1+\sum_{i}^{N/2}E(\sigma_i\sigma_0)\Big]
\label{eq:6}
\end{eqnarray}
The quantity $E(\sigma_i\sigma_0)$  is the spin-spin correlation function $S_{xx}(r_i,0)$. The use of Eq.(5) yield the following estimate:
\begin{equation}
DM_x\leq\frac{N}{2}\Bigg(1+a_1\int\limits_{b_1}^{\sqrt{N}}\frac{dr}{r}\Bigg)\Rightarrow\frac{DM_x}{N}\leq a_2+b_2\ln{N}
\label{eq:8}
\end{equation}
where $a_{1,2,}$ and $b_{1,2}$  are positive constants. The right hand side of inequality (8) increases slower than the power function (4). To reproduce the dependence (4), one has to assume that at large distances the correlation function behaves as  $S(r)\propto r^{-2+\beta}$ with the parameter $\beta=0.74$, i.e., correlations decay much slower than dictated by dipolar law (5), namely as $S(r)\propto r^{-1.26}$. Thus, Eq.(4) is at variance with the assertion on the dipolar character of the correlation function, at least in two dimensions.

This variance should be discussed in more detail. The result expressed by Eq.(5) is based on theoretical investigations of two types. First, it is often stated that such a behavior corresponds to the exact solution obtained in \cite{Sutherland1968}. This is just an extension of the results of \cite{Sutherland1968} beyond its applicability domain. In \cite{Sutherland1968}, the exact solution for the correlation functions was derived for the six vertex model with the parameter $\Delta =0$, whereas for the two-dimensional spin ice $\Delta = 1/2$. For exactly solvable models such a difference  can be critically important and can change the result qualitatively. Second, references are made to works \cite{Youngblood1980, Huse2003, Isakov2004, Henley2005} which involve the representation of the magnetization distribution function in the form
\begin{equation}
P\{{\bf M}\}\propto\exp{[s(\bf M)]}, \quad s({\bf M})=s(0)-\frac{K}{2}{\bf M}^2
\label{eq:9}
\end{equation}
where $s({\bf M})$ is the entropy of the system at given magnetization. However, the use of Eq.(9) actually implies the use of the Gaussian distribution of magnetization from the very outset. And as was mentioned above, the result expressed by Eq.(3) forces us to question the validity of this assumption (at least in two dimensions). Formally, a non-Gaussian distribution implies that the function $s({\bf M})$  has a more complicated form than that given in Eq.(9), it can include higher terms of the expansion in powers of the magnetization \cite{Ryzhkin1997} or even have no expansion at all.

Inapplicability of formulas (9) in two dimension is confirmed by the following result. If one presents the density of configurations by formula (9), the quantity $\partial{\ln{G(M)}}/\partial M^2$ will play the role of the experimental value of the coefficient K in Eq. (9). The dependence of this quantity on the number of spins in the cluster is shown in Fig.4. In this case, for noninteracting spins, one has $K\sim N^{-1}$ (see line 1). Since the total magnetization is proportional to $N$, the entropy  $s({\bf M})$  also proves to be proportional to $N$; i.e., it is an extensive quantity. On the contrary, for system with the ice rule,  $K\sim N^{-1.23}$ (see line 2) and the conditional entropy equals $s{\bf (M)}\sim N^{0.77}$, i.e., non-extensive. The nature of this non-extensivity as well as the result expressed by Eq.(3) are explained by the long-range character of correlations in spin ice. It should be emphasized that the non-extensivity is due not to the long-range character of the interaction (it is extremely short range) but exactly to the long-range spin-spin correlations.

\begin{figure}
\includegraphics[width=0.5\textwidth]{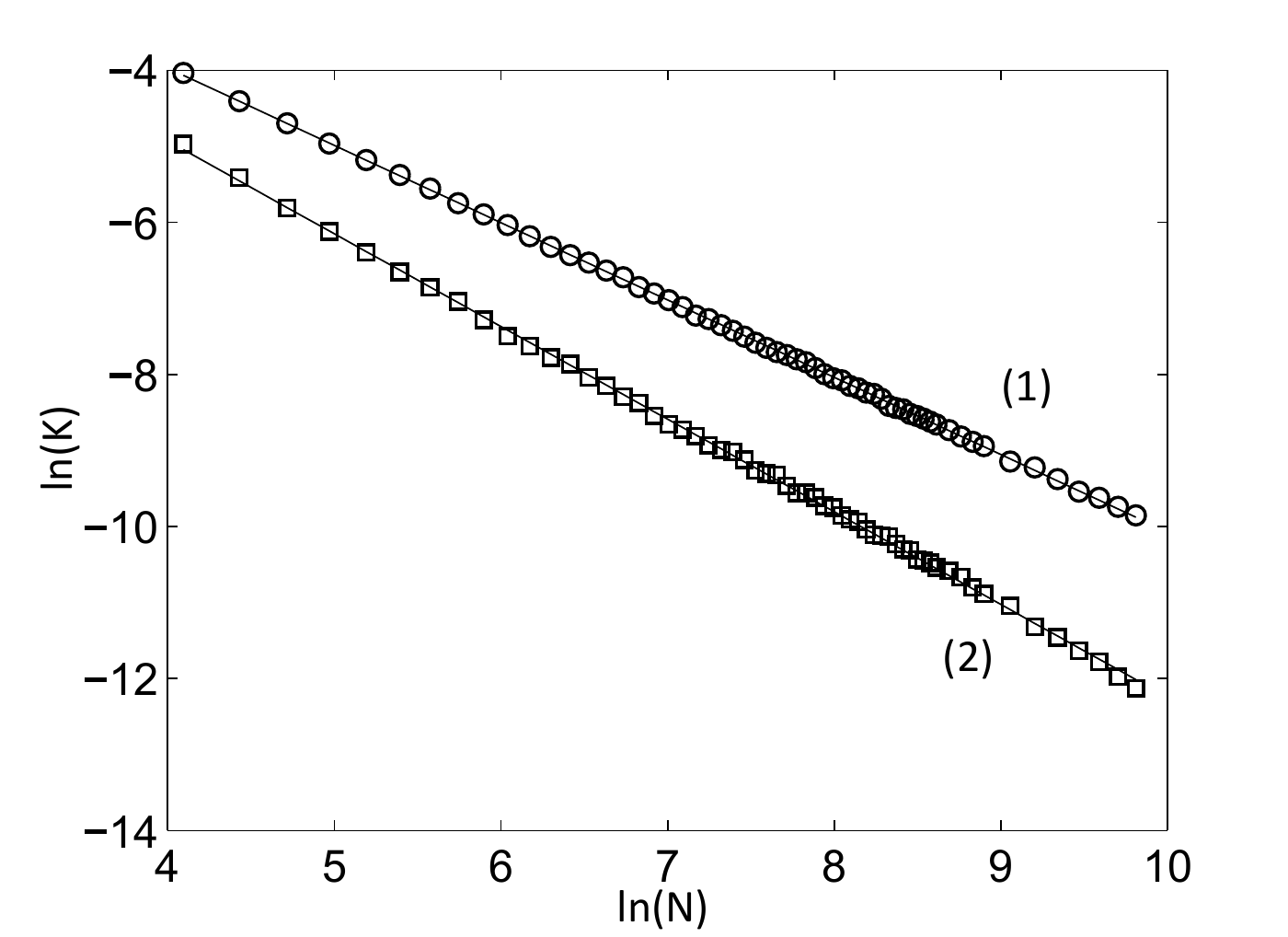}
\caption{Dependence of $K=\partial{\ln{G(M)}}/\partial M^2$ on the number of spins. This quantity defines the entropy as a function of magnetization (see Eqs.9).}
\label{fig.4}
\end{figure}

Our results also shows that the ice rule leads to broadening of the probability density function for magnetization compared to the Gaussian one for the noninteracting spins. The test of the hypothesis about Gaussian character of density distribution function by Pearson method yields the Pearson parameter  $\chi^2=3\cdot10^3$. That is greater than the quantile $\chi_{\gamma,97}^2$  for all confidence probabilties $\gamma$  (we used $100$  intervals, the number of the degrees of freedom was $k=100-2-1=97$). That means that the hypothesis of the Gaussian character of the experimental density cannot be accepted at any confidence probability. We found the experimental distribution function for the spin ice to be closest to the self-similar distribution $f(x,t)=t^{-H}p(xt^{-H})$ with the parameter $H\approx0.68$. In this case, the role of the discrete time is played by the number of spins in the system, $t \sim N$. Distribution functions of this kind appear
in the theory of anomalous diffusion (normal diffusion corresponds to $H\approx0.5$).

In conclusion, we summarize the basic results of this work. The behavior of the variance indicates correlations which decay slower than  dipolar correlations (see Fig.3 and the estimates). The conventional entropy is a nonextensive quantity (see Fig.4). The magnetization of two-dimensional spin ice does not obey the central limit theorem; i.e., the distribution of the magnetization is non-Gaussian (see Fig.4 and the results of the hypothesis testing). All these results are the consequences of long-range spin-spin correlations.
\begin{acknowledgments}
I am grateful to S. N. Molotkov and I. A. Ryzhkin for the formulation of the problem, support of the work, and discussion of the results.
\end{acknowledgments}

\bibliography{nondipole}

\end{document}